\documentclass[conference]{IEEEtran}
\IEEEoverridecommandlockouts
\usepackage{cite}
\usepackage{amsmath,amssymb,amsfonts}
\usepackage{algorithmic}
\usepackage{graphicx}
\usepackage{textcomp}
\usepackage{xcolor}
\def\BibTeX{{\rm B\kern-.05em{\sc i\kern-.025em b}\kern-.08em
    T\kern-.1667em\lower.7ex\hbox{E}\kern-.125emX}}
\begin{document}

\title{From XR Sketches to CAD: Preserving Design Intent with AI
\thanks{This work was supported by the National Science Foundation under Grant No.~2552514 and Fletcher Jones Foundation Fellowship of Dr. Trudi Di Qi.}
}

\author{\IEEEauthorblockN{Trudi Di Qi}
\IEEEauthorblockA{\textit{Fowler School of Engineering} \\
\textit{Chapman University}\\
Orange, CA, USA \\
dqi@chapman.edu}
}

\maketitle

\begin{abstract}
Extended reality (XR) allows designers to sketch shapes directly in 3D, but converting these sketches into editable, feature-based CAD requires interpreting their intent. A sketch may contain drawing errors, intended features, and design choices that remain open. Correcting every irregularity risks erasing intended features, while automatic completion may impose choices the designer has not made. We argue that AI should add precision without overriding design intent, considering both the resulting shape and how the CAD model behaves during later edits. To guide this process, we propose a conceptual framework that distinguishes errors from intentions in individual sketch properties and relates them to correction, preservation, and exploration in editable CAD. We also outline five research directions for sketch interpretation, CAD construction, and evaluation, with testable hypotheses for selected questions.
\end{abstract}


\section{Introduction}\label{sec:intro}
In product development, parametric CAD models are widely used to represent designs~\cite{popescu_2026}. A feature-based CAD model represents a part through a sequence of features, such as extrusions and holes, often built on sketched profiles. Geometric constraints and dimensions define relationships within each profile and between features. Together, they capture the model's design intent and guide how it responds to later edits~\cite{otey_2018}. These models support manufacturing, analysis, and simulation.

Extended reality (XR) allows designers to explore early ideas by sketching directly in 3D. However, freehand 3D sketches are often imprecise, limiting their direct use in downstream design tasks~\cite{yu_2021}. Converting these sketches into feature-based CAD requires precise geometry and explicit relationships between features. The challenge is to determine which properties reflect the intended design and which result from drawing errors.

\begin{figure*}[!t]
\centerline{\includegraphics[width=\textwidth]{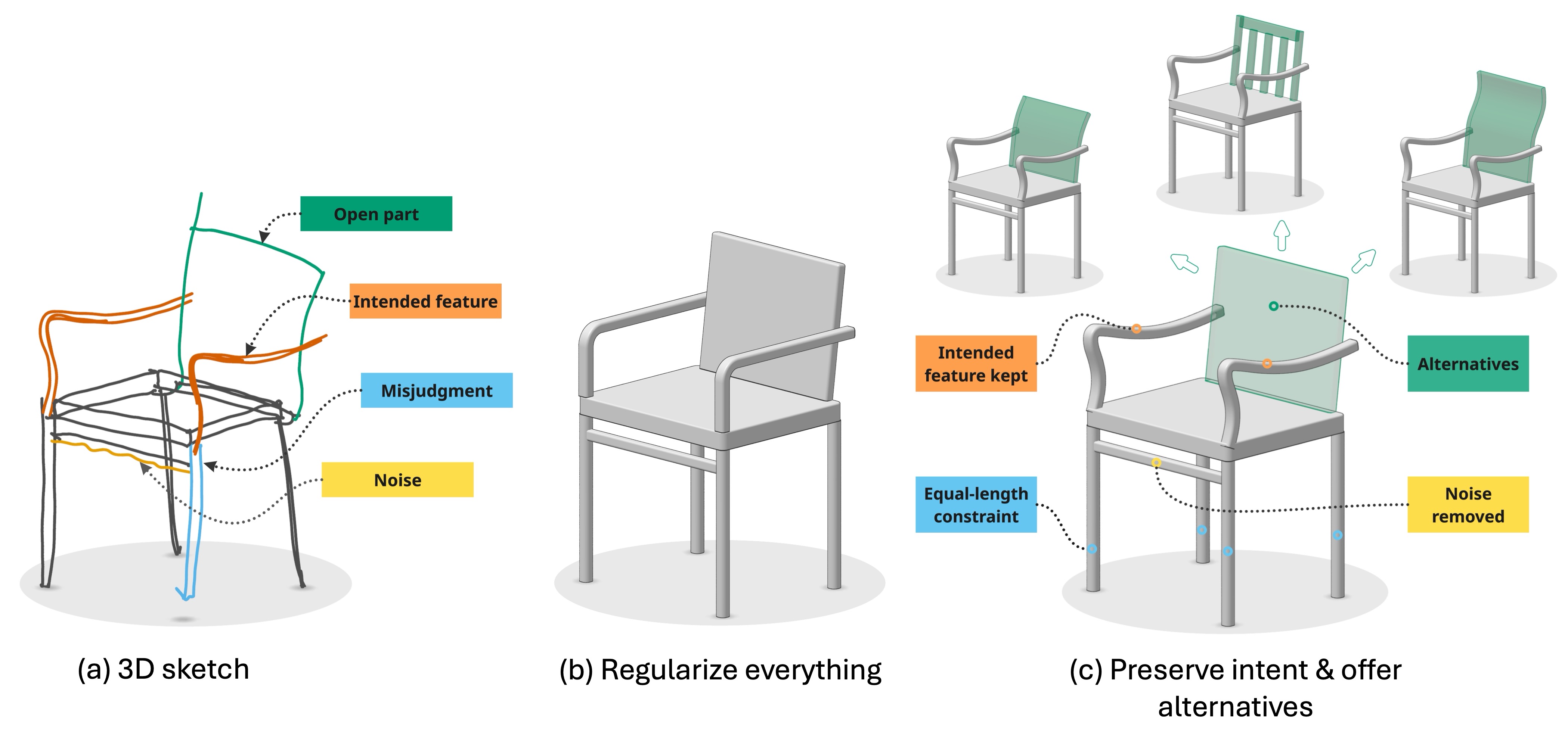}}
\caption{AI-assisted sketch-to-CAD should add precision without erasing design intent. (a) A sketch combines drawing errors (a wobbly rail and a misjudged leg length) with intentionally curved armrests and a backrest whose final form remains open. (b) Indiscriminate correction and completion remove the errors but also straighten the armrests and impose a flat backrest. (c) The proposed response removes noise, encodes equal leg lengths as a constraint, preserves editable curved armrests, and offers backrest alternatives consistent with the rest of the sketch. Preserving intent therefore concerns both the resulting shape and the relationships and choices retained in CAD. The sketch is adapted from VRChairSketch~\cite{luo_2021}; the interpretation of its strokes and all CAD models were created for illustration.}
\label{fig:teaser}
\end{figure*}

Recent research has begun to address these challenges. Generative models produce 3D shapes from VR sketches~\cite{luo_2023,chen_2025}, while other systems recognize CAD features in imprecise VR input~\cite{kukreja_2026} or convert freehand strokes into constrained CAD profiles~\cite{duan_2025}. However, producing a plausible shape or a valid CAD model does not necessarily reveal which sketch properties should be corrected, preserved, or left open for exploration. Considering the chair in Fig.~\ref{fig:teaser}a, the wobbly rail and shortened leg are drawing errors, whereas the curved armrests reflect a design choice and the sparsely drawn backrest leaves its form open. Correcting all irregularities may erase intended features, while retaining them all preserves drawing errors. Automatically completing the backrest may also impose a design choice the designer has not yet made.

We therefore argue that AI should add precision when converting 3D sketches to CAD without overriding the designer's intent. To support this vision, we first examine intent preservation in terms of both intended shape and behavior during later edits (Section~\ref{sec:intent}). We then introduce a conceptual framework that distinguishes errors from intentions in individual sketch properties and relates them to correction, preservation, and exploration in editable CAD (Section~\ref{sec:framework}). Finally, we outline five research directions for interpreting sketches, guiding CAD construction, and evaluating intent preservation, with testable hypotheses for selected questions (Section~\ref{sec:agenda}). We focus on feature-based parametric CAD, rather than freeform styling surfaces or meshes for visual content.

\section{Related Work}\label{sec:related}

\subsection{3D Sketching and Shape Modeling in XR}\label{sec:related-xr}
Sketching in XR allows designers to draw strokes directly in 3D space, but mid-air strokes are less accurate than those drawn on a physical surface~\cite{arora_2017}. Accuracy also varies with the designer's spatial ability~\cite{barreramachuca_2019b}. Assisted sketching tools aim to reduce these errors during drawing. Multiplanes adjusts strokes toward likely intended shapes~\cite{barreramachuca_2018}, while Smart3DGuides identifies common errors in freehand VR drawing and provides visual guides to reduce them~\cite{barreramachuca_2019a}. VR sketch datasets can also record timestamps for individual points~\cite{luo_2021}, providing information about the drawing process beyond the final geometry.

Geometric methods reconstruct curves and surfaces from sketch strokes. CASSIE refines individual strokes into a connected curve network and fits surface patches, balancing stroke fidelity with geometric constraints and connectivity~\cite{yu_2021}. However, some users felt that their strokes were over-neatened. For 2D input, True2Form selectively enforces regularities when reconstructing 3D curves~\cite{xu_2014}.

Learning-based approaches instead generate 3D shapes from VR sketches. Luo et al.\ produce several plausible shapes from ambiguous sketches, balancing fidelity to the input with realism and diversity~\cite{luo_2023}. The idea of keeping ambiguous input open has also been explored in earlier sketching interfaces~\cite{gross_1996}. Chen et al.\ further show that stroke order can improve shape generation~\cite{chen_2025}.

Recent XR systems also integrate sketching with generative AI. SpatialPrompt converts 3D strokes and speech into spatial constraints for mesh generation, using boundaries, proportions, and structure to guide what is retained or completed~\cite{yu_2026}. GenerativeSurface takes a different approach, generating a surface from text that the designer can sketch on~\cite{chai_2026}. While these approaches support 3D sketching and shape generation, they do not directly provide the editable feature history and constraints needed for parametric CAD.

\subsection{CAD from Sketches and Imprecise Input}\label{sec:related-cad}
Other approaches aim to construct parametric CAD directly from imprecise input. Kukreja et al.\ allow users to sculpt a voxel block in VR one feature at a time~\cite{kukreja_2026}. A neural network recognizes each sculpted feature as one of eight standard types, and a CAD kernel constructs the corresponding feature. Similarly, pARametric converts freehand strokes drawn on planes in AR into constrained profiles, after which the designer applies CAD operations~\cite{duan_2025}. Prinz et al.\ integrate a CAD kernel into a standalone VR application~\cite{prinz_2025}.

Related work has also explored CAD reconstruction from 2D drawings and existing 3D geometry. Free2CAD parses freehand drawings into CAD commands~\cite{li_2022}, while Sketch2CAD interprets each sketch as an operation on a partial model~\cite{li_2020}. Popescu and Iacob extract an intermediate representation from hand-drawn sketches and generate CAD after human validation~\cite{popescu_2026}. For 3D input, Lambourne et al.\ reconstruct editable CAD models from voxel shapes~\cite{lambourne_2022}, while Li et al.\ detect approximate symmetries in CAD models to recover design intent~\cite{li_2010}.

Learning-based methods also generate CAD sketch primitives and constraints~\cite{seff_2022}. Casey et al.\ use constraint-solver feedback to improve generated constraints and their stability during editing~\cite{casey_2025}. These methods address how design intent is represented in CAD, but a stable constraint set does not necessarily reflect what the designer intended in a freehand sketch.

\begin{table*}[!t]
\caption{Selected approaches to sketch interpretation and CAD construction. Objectives and input treatment summarize the cited work; they are not a common evaluation protocol. The vision row describes proposed targets, not demonstrated capabilities.}
\label{tab:approaches}
\centering
\footnotesize
\renewcommand{\arraystretch}{1.15}
\setlength{\tabcolsep}{4pt}
\begin{tabular}{@{}p{0.92in}p{1.33in}p{1.17in}p{1.25in}p{2.03in}@{}}
\hline
\raggedright\textbf{Approach} & \raggedright\textbf{Input} & \raggedright\textbf{Output} & \raggedright\textbf{Main objective} & \raggedright\textbf{How input guides the result} \tabularnewline
\hline
\raggedright CASSIE~\cite{yu_2021} & \raggedright 3D strokes (VR) & \raggedright Curve network, surface patches & \raggedright Clean, connected curves & \raggedright Selects constraints while balancing stroke fidelity \tabularnewline
\raggedright Luo et al.~\cite{luo_2023} & \raggedright VR sketch & \raggedright 3D shape & \raggedright Fidelity to sketch, realism, diversity & \raggedright Generates alternative shapes conditioned on the sketch \tabularnewline
\raggedright SpatialPrompt~\cite{yu_2026} & \raggedright 3D strokes and speech (XR) & \raggedright Generated mesh & \raggedright Spatially guided generation & \raggedright Sketch and speech guide structure and completion \tabularnewline
\raggedright Kukreja et al.~\cite{kukreja_2026} & \raggedright Voxel sculpting (VR), one feature at a time & \raggedright Parametric CAD & \raggedright CAD feature recognition & \raggedright Classifies sculpted input into a predefined feature set \tabularnewline
\raggedright pARametric~\cite{duan_2025} & \raggedright Strokes on planes (AR) & \raggedright Constrained profiles and CAD model & \raggedright In situ parametric modeling & \raggedright Infers profile constraints; user applies CAD operations \tabularnewline
\raggedright Free2CAD~\cite{li_2022} & \raggedright Freehand 2D drawing & \raggedright CAD command sequence & \raggedright CAD reconstruction from drawings & \raggedright Maps stroke groups to CAD commands \tabularnewline
\raggedright Vitruvion~\cite{seff_2022} & \raggedright Partial sketches or sketch images (2D); also unconditioned generation & \raggedright CAD sketch primitives and constraints & \raggedright Parametric sketch generation & \raggedright Learns geometry and constraint distributions \tabularnewline
\raggedright Casey et al.~\cite{casey_2025} & \raggedright CAD sketch primitives (2D) & \raggedright CAD sketch constraints & \raggedright Well-constrained sketches and stable edits & \raggedright Uses solver feedback to guide constraint generation \tabularnewline
\raggedright This paper (vision) & \raggedright Whole-object 3D sketch & \raggedright Editable feature-based CAD (target) & \raggedright Intended shape and editing behavior & \raggedright Distinguish errors, intended features, and open design choices (proposed) \tabularnewline
\hline
\end{tabular}
\end{table*}

\subsection{Positioning of This Work}\label{sec:positioning}
A recent survey of deep sketch-based 3D modeling does not cover immersive 3D sketching~\cite{tono_2026}. The studies reviewed above address different aspects of this problem, from correcting drawing errors and handling ambiguity to reconstructing geometry and recovering CAD constraints. However, these objectives are generally addressed separately, with less attention to how individual properties of a 3D sketch reflect the designer's intent and affect subsequent CAD edits. Some relationships also become clearer when considering the finished sketch, such as whether two armrests share a similar curve (Fig.~\ref{fig:teaser}a). We therefore focus on finished XR sketches of whole objects, examining which properties should be corrected, preserved, or left open, and how these decisions affect both the resulting shape and its editing behavior.

Our framework therefore considers errors and intentions at the level of individual sketch properties, allowing both to coexist within the same part. Table~\ref{tab:approaches} compares selected approaches in terms of their objectives and treatment of input. Section~\ref{sec:intent} further discusses the requirements for preserving intent in both geometry and editing.

\section{Design Intent in Sketch-to-CAD}\label{sec:intent}

\subsection{Limits of Geometric Fidelity}\label{sec:fidelity}

Freehand 3D sketches often contain irregularities, such as wobbly strokes, uneven lengths, and curves that depart from regular shapes. Methods for improving sketch precision generally balance two goals: staying close to the drawn strokes and adjusting them toward more regular or typical forms. Sketch-conditioned generation considers stroke fidelity alongside realism for the object category~\cite{luo_2023}. CASSIE balances fidelity to individual strokes with regular relationships between curves~\cite{yu_2021}. Other methods favor symmetry~\cite{zhou_2025} or infer standard CAD features~\cite{kukreja_2026} and geometric constraints~\cite{seff_2022,duan_2025}. While these approaches address imprecision in different ways, they do not necessarily distinguish intentional features from drawing errors.

This distinction becomes clear when a sketch contains both types of irregularity. In Fig.~\ref{fig:teaser}a, the wobble in the front rail is a drawing error, whereas the armrests are intentionally curved. Strict fidelity to the strokes retains the wobble, while regularization may remove it at the cost of straightening the armrests (Fig.~\ref{fig:teaser}b). A suitable result should correct the rail while preserving the armrest curvature, which requires interpreting each irregularity rather than treating them all alike.

Evaluation must also distinguish geometric similarity from intent preservation. Common metrics, such as Chamfer distance, measure geometric differences between generated and reference shapes but do not directly capture the designer's intent~\cite{tono_2026}. For sketch-to-CAD, evaluation should examine whether the resulting model retains features the designer intended to include and avoids prematurely fixing the form of parts the designer wanted to leave open.

\subsection{Intent in Geometry and Editing}\label{sec:intent-editing}
In parametric CAD, design intent is closely tied to how a model responds to later edits~\cite{otey_2018}. Features, constraints, and parameters encode relationships that guide this behavior. For example, geometric constraints can maintain parallelism or equal dimensions as the model changes~\cite{casey_2025}. Two models may therefore have the same shape but respond differently to edits, depending on how they are constructed and constrained.

We connect the designer's intent in an XR sketch to how the resulting CAD model behaves during editing. AI-assisted sketch-to-CAD should therefore satisfy two requirements: the model should reflect the intended shape, and its constraints and parameters should preserve intended relationships during later edits. This requires distinguishing errors to correct from intentions to preserve, including both chosen features and aspects of form the designer wants to leave open. When the sketch provides insufficient evidence, the AI may need to ask the designer before making a change. Section~\ref{sec:framework} develops this distinction into a framework for handling different sketch properties in editable CAD.

\begin{table*}[!t]
\caption{Errors and intentions in a 3D sketch, with possible cues and CAD responses illustrated by the chair in Fig.~\ref{fig:teaser}. The cues are proposed hypotheses, not validated findings.}
\label{tab:framework}
\centering
\footnotesize
\renewcommand{\arraystretch}{1.15}
\setlength{\tabcolsep}{4pt}
\begin{tabular}{@{}p{1.05in}p{1.35in}p{1.35in}p{1.35in}p{1.6in}@{}}
\hline
\raggedright\textbf{Form and kind} & \raggedright\textbf{Chair example} & \raggedright\textbf{Possible geometric cues} & \raggedright\textbf{Other cues} & \raggedright\textbf{CAD response} \tabularnewline
\hline
\raggedright Noise (Error) & \raggedright Wobbly rail & \raggedright Small, local fluctuations & \raggedright Quick, unrevised strokes & \raggedright Remove unintended variations \tabularnewline
\raggedright Misjudgment (Error) & \raggedright One leg too short & \raggedright Nearly equal lengths & \raggedright Parts drawn as a set & \raggedright Recover intended dimensions or relationships \tabularnewline
\raggedright Intended feature (Intention) & \raggedright Intentionally curved armrests & \raggedright Consistent curvature across strokes & \raggedright Repeated or redrawn strokes; mark to retain & \raggedright Preserve defining form and editability \tabularnewline
\raggedright Open part (Intention) & \raggedright Backrest drawn with less detail & \raggedright Strokes admit several forms & \raggedright Few strokes without refinement; mark to explore & \raggedright Offer alternatives for designer selection \tabularnewline
\hline
\end{tabular}
\end{table*}

\section{A Framework for Intent Preservation}\label{sec:framework}
To guide how AI interprets XR sketches, we distinguish between \textbf{errors}, or differences between what the designer meant and what was drawn, and \textbf{intentions}, including chosen features and design decisions intentionally left open. This distinction applies to individual properties, such as lengths, curves, and part shapes, allowing errors and intentions to coexist within the same part. The framework is a conceptual proposal informed by prior work and the examples below, not an empirically validated classification.

\subsection{Errors, Intended Features, and Open Parts}\label{sec:forms}
Among errors, \textbf{noise} refers to unintended variations introduced during drawing or tracking. A smooth stroke may wobble, as with the chair rail in Fig.~\ref{fig:teaser}a, while strokes meant to meet may leave a small gap or overlap. Such inaccuracies are common challenges in mid-air drawing~\cite{arora_2017,barreramachuca_2019a}. A \textbf{misjudgment} occurs when the designer draws a specific property or relationship inaccurately, such as one chair leg being shorter when all legs were intended to have equal lengths (Fig.~\ref{fig:teaser}a). Spatial ability also affects accuracy in immersive drawing~\cite{barreramachuca_2019b}. Nearly equal lengths or nearly parallel strokes may indicate this type of error, while techniques for detecting approximate symmetry in 3D geometry could help identify candidate relationships~\cite{mitra_2006}.

Among intentions, an \textbf{intended feature} is a property the designer chooses to retain. In Fig.~\ref{fig:teaser}a, both armrests follow a similar curve, suggesting that their curvature is intentional. In contrast, an \textbf{open part} represents a design choice the designer has not yet settled. The backrest sketch indicates height and width but leaves room for different forms, such as flat panels or slats. Its slightly curved left edge may also suggest a curved surface without defining its final form. Sparse strokes can support several interpretations~\cite{luo_2023}, but they do not necessarily indicate whether the designer left the form open or simply drew an incomplete version of an already decided design. Table~\ref{tab:framework} summarizes these four categories and their proposed CAD responses.

\subsection{Evidence and Uncertainty}\label{sec:evidence}
Geometry alone cannot reliably reveal intent: a small deviation may be intentional, while a large one may be an error. Drawing history could provide additional clues. CASSIE's authors suggest adapting neatening based on stroke speed or pressure~\cite{yu_2021}, and stroke order has been shown to improve shape generation~\cite{chen_2025}. These findings motivate examining drawing history but do not demonstrate that it can distinguish errors from intentions. Designers could also explicitly mark features to retain or parts they wish to explore. Popescu and Iacob distinguish visible, inferred, assumed, and missing information in sketch interpretations, which are validated by a person before CAD generation~\cite{popescu_2026}. When evidence is insufficient, the AI may ask for clarification or defer a change. The choice between acting, asking, and deferring should depend on both the uncertainty and potential impact of an incorrect interpretation.

\subsection{Implications for Editable CAD}\label{sec:consequences}
Once identified, \textbf{noise} can be removed from the generated geometry, with a record of the correction retained for inspection. \textbf{Misjudgments}, on the other hand, require recovering the intended dimensions or relationships rather than simply adjusting the geometry. For example, Fig.~\ref{fig:teaser}c uses an equal-length constraint to maintain consistent leg lengths during later edits. Encoding this relationship connects geometric correction with the editing behavior associated with CAD design intent~\cite{otey_2018}.

For an \textbf{intended feature}, the CAD model should preserve its defining form while keeping it editable, rather than fixing all its parameters. The armrests in Fig.~\ref{fig:teaser}a illustrate how errors and intentions can coexist: both follow a similar curve but differ slightly from each other. If the designer intended them to match, these differences can be corrected without straightening the curves. Fig.~\ref{fig:teaser}c shows one possible solution, using a shared, editable curved profile for both armrests.

An \textbf{open part} should not be assigned a final form before the designer decides. Instead, the AI could offer alternatives that fit the drawn strokes and remain consistent with the proportions and structure of the rest of the sketch. If the designer already has a particular form in mind but has not fully drawn it, assistance should focus on completing that form rather than treating it as open for exploration. In Fig.~\ref{fig:teaser}c, the backrest alternatives explore different forms within the chair's existing design. Subtle cues, such as the slightly curved backrest edge, may guide these options without fixing the outcome. The designer then chooses whether to adopt any alternative in the CAD model. These distinctions motivate the research directions in Section~\ref{sec:agenda}.

\section{Research Agenda}\label{sec:agenda}
Sections~\ref{sec:intent} and~\ref{sec:framework} identify two requirements for sketch-to-CAD: preserving intended form and supporting expected editing behavior. Meeting both requires evidence to distinguish drawing errors, intentional features, and design choices left open for exploration. We outline \textbf{five research directions} for interpreting these properties, constructing editable CAD, and evaluating the results.

\subsection{Linking Sketch Properties to CAD}\label{sec:linking}
Linking CAD elements back to the original sketch could help designers understand and revise AI interpretations. Existing methods provide useful starting points. Free2CAD groups drawing strokes and maps each group to a CAD command~\cite{li_2022}, while Sketch2CAD interprets new sketches as operations on a partial model~\cite{li_2020}. Lambourne et al.\ recover profiles and extrusions from voxels and construct constrained sketches~\cite{lambourne_2022}. Kukreja et al.\ support feature-based CAD construction in VR~\cite{kukreja_2026}.

An open challenge is to retain these links as sketch properties are translated into CAD features, constraints, and parameters. The links should indicate which properties were corrected, preserved, or left open, including relationships that span multiple strokes and CAD elements. This would help designers identify unwanted interpretations and understand which parts of the model would be affected by a revision.

\subsection{Distinguishing Errors from Intentions}\label{sec:distinguishing}
Existing geometric and generative methods offer potential cues for distinguishing errors from intentions. Approximate symmetry detection measures deviations from symmetry~\cite{mitra_2006}, while geometric relation discovery identifies relationships such as parallelism and equal size across fitted primitives~\cite{li_2011}. Generative models estimate likely CAD constraints~\cite{seff_2022}, but a plausible constraint does not necessarily reflect the designer's intent. Drawing history provides another source of information: VR sketch datasets record timestamps~\cite{luo_2021}, and stroke order has been shown to improve shape generation~\cite{chen_2025}.

Whether these signals can reliably distinguish errors from intentions remains an open question. We propose \textbf{three hypotheses} that can be tested against designers' own accounts of what they intended to draw.

\textbf{H1}. Irregularities such as unequal spacing or asymmetry are more likely to be intentional when repeated consistently across related strokes than when comparable deviations occur only once.

\textbf{H2}. Drawing history, including speed, pauses, and revisions, provides a more reliable basis than geometry alone for distinguishing intentionally open design choices from incompletely drawn but already decided forms.

\textbf{H3}. When a sketch is ambiguous, asking the designer before making a change reduces the combined effort of clarification and correcting unwanted results, compared with making changes without asking.

These hypotheses should be tested across different drawing styles and motor abilities. How to request clarification and present alternatives in the headset without disrupting the XR workflow remains an open question for 3D interaction.

\subsection{Preserving Intentions and Supporting Alternatives}\label{sec:preserving}
Existing methods provide technical starting points for the CAD responses discussed in Section~\ref{sec:consequences}. Geometric fitting and relation discovery can help correct inaccuracies~\cite{li_2011}, while Vitruvion completes partial sketches with primitives and constraints~\cite{seff_2022}. Casey et al.\ use constraint-solver feedback to improve generated constraint sets~\cite{casey_2025}, offering a basis for preserving intended relationships during editing. However, solver feedback alone cannot confirm whether a constraint matches the designer's intent. Sketch-conditioned generators can also produce multiple shapes from a single sketch~\cite{luo_2023}, suggesting a way to explore open parts while respecting the strokes, proportions, and structure of the existing design.

The remaining challenge is to control how AI modifies individual properties, even within the same part. Rather than refining all geometry uniformly, the AI should correct errors, preserve intended features, and leave unresolved design choices open for exploration. For example, differences between the armrests could be corrected without losing their shared curved form (Fig.~\ref{fig:teaser}c). Evaluation should therefore examine whether these operations preserve intended properties and editing behavior, not simply whether they produce valid CAD geometry.

\subsection{Data with Designer-Provided Intent}\label{sec:data}
Existing resources include VR sketches paired with meshes~\cite{luo_2021,chen_2025}, CAD model collections~\cite{koch_2019}, and sketch-and-extrude construction sequences~\cite{willis_2021}. To study intent preservation, a dataset would need 3D sketches paired with editable CAD models that reflect the designers' intentions. Beyond geometry, it should capture drawing history and designer-provided or validated annotations of errors, intended features and relationships, and intentionally open parts at the level of individual sketch properties. These annotations should distinguish original intentions from decisions made later in the design process, which cannot be determined from the final CAD model alone. Collecting drawing and motion data would require informed consent.

\subsection{Measuring Intent Preservation}\label{sec:measuring}
Common metrics, such as shape similarity and feature-recognition accuracy~\cite{tono_2026,kukreja_2026}, do not directly indicate whether the designer's intent has been preserved. Barrera Machuca et al.\ review evaluation practices for immersive sketching~\cite{barreramachuca_2023}. For sketch-to-CAD, we propose three complementary areas of assessment that connect the original sketch to the resulting editable model.

First, \textbf{intent preservation} should measure how often intended properties are retained and known errors are corrected. It should also identify unintended changes to chosen features and open parts completed without the designer's approval. For generated alternatives, evaluation should consider whether they explore unresolved aspects while remaining consistent with properties already specified in the sketch.

Second, \textbf{constraint recovery} should evaluate whether CAD dimensions and constraints capture intended relationships, maintain them during relevant edits, and allow the changes the designer expects to make. Because different constraint sets can support equivalent editing behavior, exact agreement with a single reference construction should not be required.

Third, \textbf{correction effort} should measure the work required to repair unwanted AI changes, separately from continued design development. The effort involved in answering AI clarification questions should also be included.

\subsection{Limits and Trade-offs}\label{sec:limits}
These research directions focus on feature-based CAD. For freeform surfaces, intent may be expressed primarily through curvature and continuity, requiring different representations and evaluation measures. Completing a sketch too early may limit exploration, while assumptions about typical shapes or unreliable cues may lead to unwanted changes. Design intent can also evolve, so earlier AI interpretations should remain revisable. Whether automated correction changes how carefully designers draw or supports those with limited motor control remains an open research question.

\section{Conclusion}\label{sec:conclusion}
Converting an XR sketch into editable CAD requires more than correcting its geometry. Similar-looking irregularities may reflect drawing errors or intended features, while sparse regions may represent choices the designer has not yet made. Our proposed framework connects these interpretations to possible CAD responses, considering both the resulting shape and how the model behaves during later edits. The research agenda identifies ways to investigate these distinctions and evaluate the results with designer input. Ultimately, AI assistance should correct errors without changing chosen features, introducing unwanted constraints, or closing off design options too early.

\section*{Acknowledgment}
The author used Claude (Anthropic) to assist with generating Python scripts that built and rendered the CAD models in Fig.~\ref{fig:teaser}, as well as LaTeX syntax and formatting, and ChatGPT (OpenAI) and Grammarly to assist with grammar checking and language clarity.

\bibliographystyle{IEEEtran}
\bibliography{references}

\end{document}